\documentclass[]{raa}            
\usepackage{graphicx,times}
\usepackage{float}
\usepackage{natbib}
\usepackage{wrapfig}
\providecommand{\bm}[1]{\mbox{\boldmath$#1$\unboldmath}}

\begin{document}

   \title{CO observation of the Galactic bubble N4
}

   \volnopage{Vol.13 (2013) No.8, 921--934}
   \setcounter{page}{1}

   \author{Jun-Yu Li
      \inst{1,2}
   \and Zhi-Bo Jiang
      \inst{1}
    \and Yao Liu
      \inst{1}
     \and Yuan Wang
     \inst{1}
  }

   \institute{Purple Mountain Observatory, Chinese Academy of Sciences,
             Nanjing 210008, China; {\it lijunyu@pmo.ac.cn}\\
        \and
             University of Chinese Academy of Sciences,
             Beijing 100049, China\
             \vs \no
}

\abstract{ We presented a study on the Galactic bubble N4 using the 13.7 m millimeter telescope of Purple Mountain Observatory at the Qinghai Station. N4 is one of the science demonstration regions for the Milky Way Imaging Scroll Painting (WMISP). Simultaneous observations of $^{12}$CO (J = 1$-$0), $^{13}$CO (J = 1$-$0) and C$^{18}$O (J = 1$-$0) line emission towards N4 were carried out. We analyzed the spectral profile and the distribution of the molecular gas. Morphologically, the CO emissions correlate well with Spitzer IRAC 8.0 $\mu$m emission. The channel map and velocity-position diagram shows that N4 is more likely an inclined expanding ring than a spherical bubble. We calculated the physical parameters of N4 including the mass, size, column density and optical depth. Some massive star candidates were discovered in the region of N4 using (J, J$-$H) color-magnitude diagram. We found an energy source candidate for the expansion of N4, a massive star with a mass of ${\sim} 15\,M_{\odot}$ and an age of $\sim$ 1 Myr. There exists infall motion signature in N4, which can be a good candidate of infall area. Combined mm and infrared data, we think there may exists triggered star formation in N4.
}

   \authorrunning{Jun-Yu Li et al.}            
   \titlerunning{CO observation of the Galactic bubble N4}  
   \maketitle


%
%
\section{Introduction}           
\label{sect:intro}
The Galactic bubble N4 (L = 11.892$^\circ$ and B = 0.748$^\circ$) was first revealed in the Galactic Legacy Infrared Mid-Plane Survey Extraordinaire (GLIMPSE; Benjamin et al. 2003; Churchwell et al. 2009) by the Spitzer telescope. N4 is surrounded by a shell of collected material (Deharveng et al. 2010). The distance of N4 (d = 3.2 kpc) is determined from the radial velocity of  overlapping H{\scriptsize II} region (Churchwell et al. 2006). Since the probability of chance alignments of H{\scriptsize II} region with a bubble is quite small ($<$ 1$\%$) (Churchwell et al. 2006), we assume that the bubble lie at the same distance as the H{\scriptsize II} region. The bubble at the far kinematic distance (13.4 kpc) would be so far away that only the near distance is a realistic choice, therefore in this paper the near distance, i.e., 3.2 kpc, is adopted as the distance of N4.\\
\indent The strong winds and radiation fields from massive stars clear out large cavities in the interstellar medium (ISM). These cavities (or bubbles) impact their surrounding molecular clouds and may influence the formation of stars therein (Beaumont $\&$ Williams 2010). At the boundaries of these bubbles lie the material displaced by the propagating wind or ionization front. Such a dynamic process, which alters the physical environment of molecular clouds, might trigger the formation of a new generation of stars.\\
\indent The study of the bubbles can give some information about the stellar winds that produce them and the structure and physical properties of the ambient ISM into which they are expanding. Thus the distribution of dense molecular material can reflect the interaction between the bubble and its surrounding molecular clouds. It is speculated that gravitational collapse might be triggered by the expanding bubble when the leading shock front overruns and compresses a preexisting molecular cloud, resulting in star formation. In theory, stars of all masses can form in the shell around H{\scriptsize II} region. Small-scale gravitational instabilities (for example Jeans instabilities) can lead to the formation of low-mass stars; large-scale gravitational instabilities along the collected shell can lead to the formation of massive fragments, potential sites of massive star formation; this is called the ``collect $\&$ collapse'' process (Elmegreen $\&$ Lada 1977). \\
\indent In order to study the interaction between the bubble and ambient molecular clouds, we observed the molecular clouds associated with N4. So far there is little information about the molecular clouds around N4. While the infrared and sub-mm dust continuum imaging can easily pick up the bubble morphologically, molecular line observation is crucial for searching for kinematic evidence of interactions between  H{\scriptsize II} regions and molecular clouds and testing the collect and collapse process. Therefore observations of CO ($^{12}$CO , $^{13}$CO, C$^{18}$O) line emissions towards N4 were done simultaneously. \\
\indent The observations and data reduction are described in Section 2; the results and data analysis are presented in Section 3; Section 4 is the discussion and Section 5 is the conclusions.


\section{Observations and Data reduction}
\label{sect:Obs}

The observation of N4 was carried out from 2011 December 24 to December 26 as a part of Multi-Line Galactic Plane Survey in CO and its Isotopic Transitions, also called the Milky Way Imaging Scroll Painting (WMISP). It is a project dedicated for the large-scale survey of dense molecular gas along the northern Galactic Plane. Regions within the Galactic range of $-$5.25$^\circ$ $\leq$ B $\leq$ 5.25$^\circ$, $-$10.25$^\circ$ $\leq$ L $\leq$ 250.25$^\circ$ and some other interested areas are divided into $\sim$11000 cells unit in (L, B) grid, each with a size of 30$^{'}$ $\times$ 30$^{'}$. The Superconducting Spectroscopic Array Receiver (SSAR), a 3 $\times$ 3 beam superconducting focal plane array mounted on Delingha 13.7 m millimeter telescope is used (Shan et al. 2012;  Zuo et al. 2011). The array receiver was made by Superconductor-Insulator-Superconductor (SIS) mixers with the sideband separation scheme. A specific LO frequency was selected so that the upper sideband is centered at the  $^{12}$CO (1$-$0) line and the lower sideband covers the  $^{13}$CO (1$-$0) and C$^{18}$O (1$-$0) lines which enables us to make simultaneous detection of the CO isotopic lines. The back end is a Fast Fourier Transform Spectrometer (FFTS) of 16384 channels with a bandwidth of 1000 MHz and an effective spectral resolution of 61.0~KHz (0.16~km s$^{-1}$). The size of the main beam is about 52$^{''}$ at 115 GHz and the main beam efficiency $\eta_{mb}$ is 0.46 at upper sideband, 0.43 at lower sideband. The absolute pointing uncertainty was estimated to be $\sim$ 5$^{''}$ from continuum observations of planets. Using On-The-Fly (OTF) scanning, with the scanning rate of 75$^{''}$per second and the total integration time of 54 s one pixel, a large region ($-$1$^\circ$ $\leq$ B $\leq$ 1$^\circ$, 12.5$^\circ$ $\leq$ L $\leq$ 16$^\circ$) has been observed, but in this work we take a region ($\sim$ 21.6$^{'}$ $\times$ 21.6$^{'}$) as a main target. During the observation, we took spectra of the standard source W51D every two hours to check whether the telescope and the other devices worked properly or not. We used the standard chopper wheel calibration technique to measure antenna temperature $T_{A}^{*}$ corrected for atmospheric absorption. The final data was recorded in brightness temperature scale of $T_{R}^{*}$ (K). The system temperatures were about 250 K$-$350 K during the observation. \\
\indent The spectral data were reduced and analyzed with the GILDAS/CLASS\footnote[1]{http://iram.fr/IRAMFR/GILDAS/} package. As a preliminary step of data reduction, all spectra of 9 beams were individually inspected and those with either strongly distorted baselines or abnormal rms noise levels were discarded. We combined the spectra of the same position and made linear baseline subtractions to the adopted spectra. The RMS is about 0.48 K for  $^{12}$CO and 0.27 K for $^{13}$CO and C$^{18}$O.


\section{RESULTS and Data ANALYSIS}
\label{sect:result}

\subsection{The morphologies of the $^{12}$CO (1$-$0), $^{13}$CO (1$-$0) and C$^{18}$O (1$-$0) emission}
Figure 1 shows the velocity-integrated CO emission, which we call $W_{CO}$(K km s$^{-1}$).  In the maps of $W_{CO}$, the red circle indicates the position of N4 region ($\sim$ 7.2$^{'}$ $\times$ 7.2$^{'}$), we see molecular CO emissions of N4 are almost circular. The most striking feature of CO data is the rarity of emission toward the center of N4. The maps of $W_{CO}$ also show clumpy structures along the shell of N4. It suggests that the physical conditions (temperature, density) are not uniform in N4. The emission of $^{12}$CO and $^{13}$CO is strong, while the C$^{18}$O emission is week. We take contour levels starting at 30$\%$ ($\sim$36 K km s$^{-1}$) of the peak $^{12}$CO integrated intensity, at a step $\sim$12 K km s$^{-1}$ in the first panel; contour levels start at 30$\%$ ($\sim$12 K km s$^{-1}$) of the peak $^{13}$CO integrated intensity, at a step $\sim$ 4 K km s$^{-1}$ in the second panel; contour levels start at 5$\sigma$ ($\sigma$$\sim$0.34 K km s$^{-1}$) of the C$^{18}$O integrated intensity, at a step $\sim$ 0.57 K km s$^{-1}$ in the third panel. As one steps through the velocity space and the spectra line of CO emission, there are several components in the line of our sight, while the emission from N4 has a velocity range of 21 km s$^{-1}$ $<$ $V_{LSR}$ $<$ 28 km s$^{-1}$. Thus the integrated velocity range of the three panels is all from 21 km s$^{-1}$ to 28 km s$^{-1}$. 

\begin{figure}[!htbp]
\centering
\includegraphics[width=76mm]{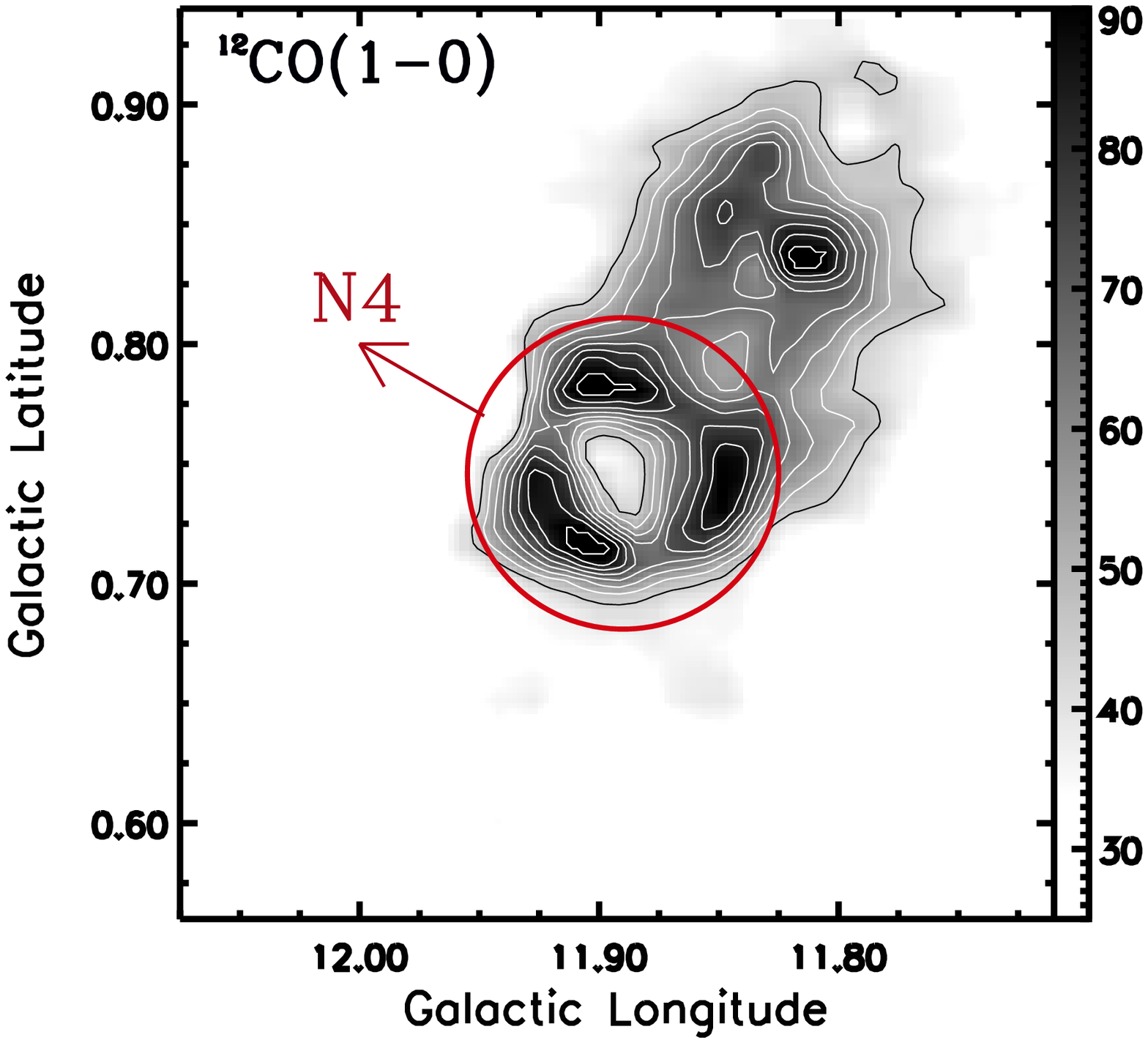}\\
\includegraphics[width=76mm]{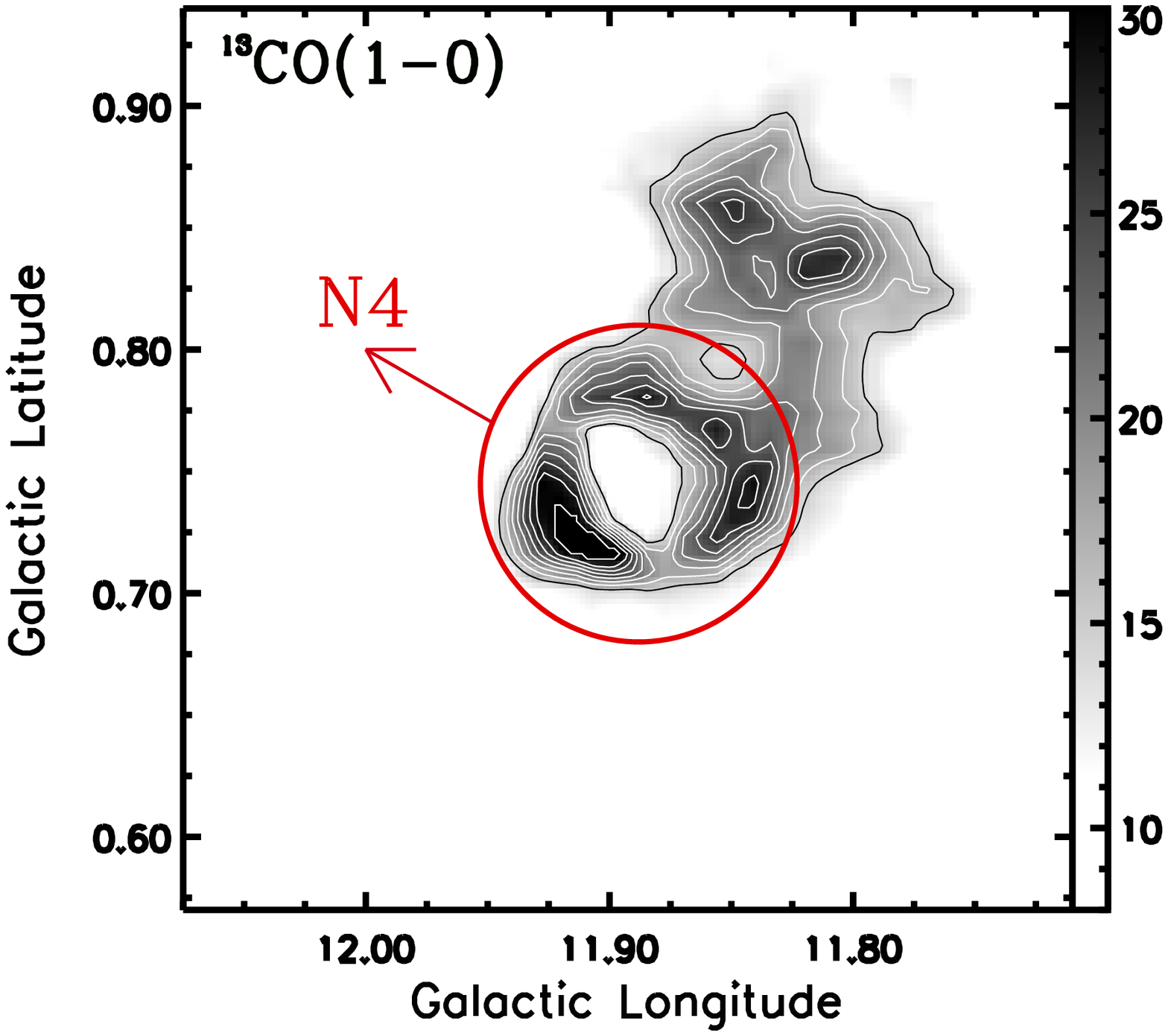}\\
\includegraphics[width=76mm]{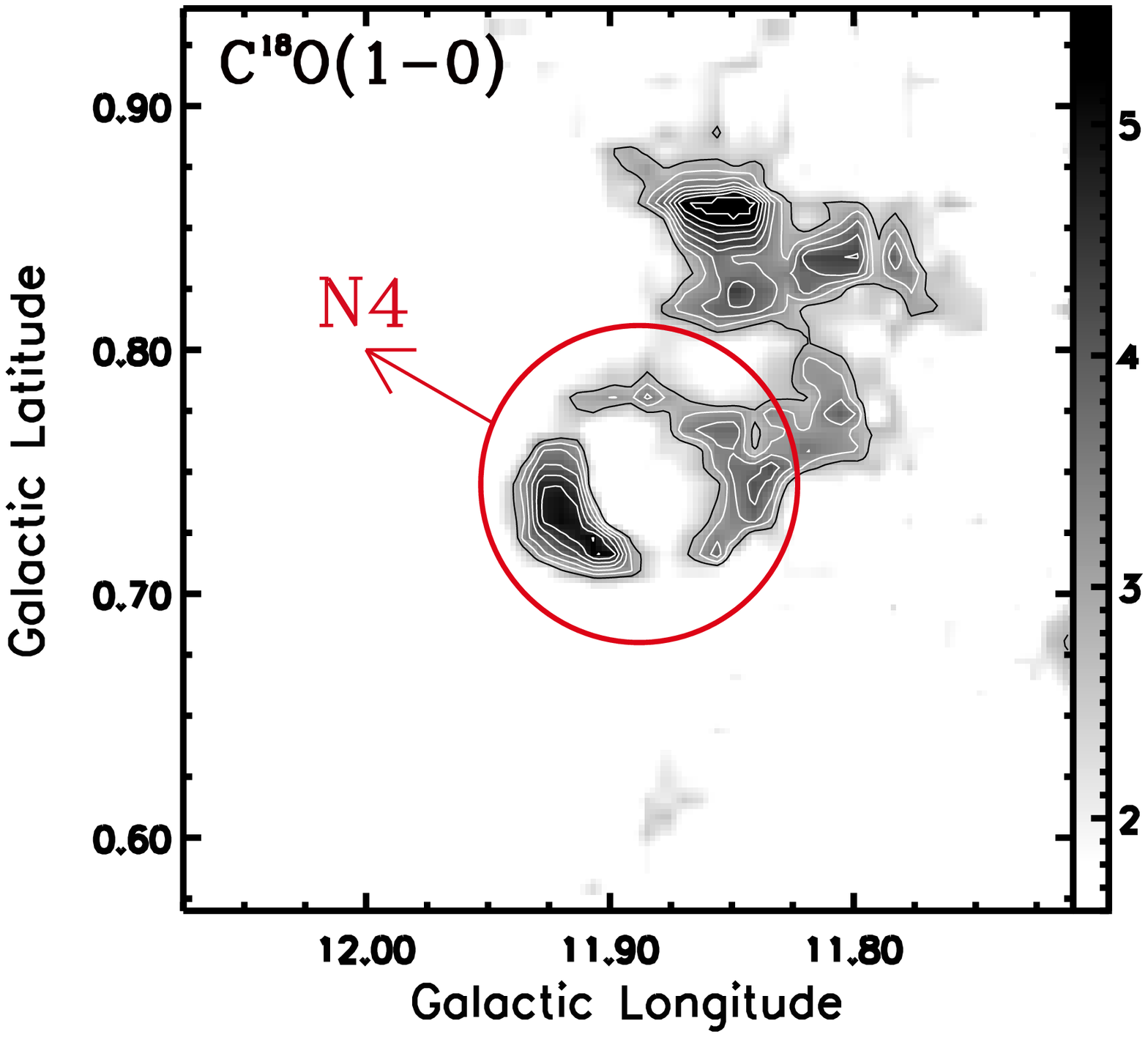}
\caption{First panel: the integrated intensity map of $^{12}$CO. Contour levels start at 30$\%$ ($\sim$ 36 K km s$^{-1}$) of the peak $^{12}$CO integrated intensity, at a step $\sim$ 12 K km s$^{-1}$. Second panel: the integrated intensity map of $^{13}$CO. Contour levels start at 30$\%$ ($\sim$ 12 K km s$^{-1}$) of the peak $^{13}$CO integrated intensity, at a step $\sim$ 4 K km s$^{-1}$. Third panel: the integrated intensity map of C$^{18}$O. Contour levels start at 5$\sigma$ ($\sigma$ $\sim$ 0.34 K km s$^{-1}$) of the C$^{18}$O integrated intensity, at a step $\sim$ 0.55 K km s$^{-1}$. The integrated velocity range is all from 21 km s$^{-1}$ to 28 km s$^{-1}$. The red circle indicates the position of N4. The unit of the color bar is in K km s$^{-1}$.}
   \label{Fig1}
\end{figure}
\subsection{The infrared dust and CO distribution of N4} 
In Fig. 2, we present overlay maps of the $^{12}$CO, $^{13}$CO and C$^{18}$O data
on the infrared data. The data of Spitzer IRAC 8.0 $\mu$m and 13.7 m CO emissions of N4 show that the H{\scriptsize II} region is surrounded by a remarkably complete ring. According to Churchwell et al. (2006), bubbles found in GLIMPSE are generally detectable in all four IRAC bands but tend to be brighter at longer wavelengths. Most bubbles are larger and brighter at 8.0 $\mu$m than at the shorter wavelength. This is also the case of N4. The 8.0 $\mu$m emission is faint or absent inside the bubble, and bright along the shell that defines the bubble, and generally extends well beyond the bubble shell boundary. Churchwell et al. (2006) postulated that the 8.0 $\mu$m band emission is probably dominated by the 7.7 $\mu$m and 8.6 $\mu$m features attributed to polycyclic aromatic hydrocarbons (PAHs). The fact that 8.0 $\mu$m emission appears to be essentially absent in the central regions of the bubbles (the front and back sides of the shell that defines the bubble contribute some emission) implies that PAHs are either easily destroyed in the hard radiation field of the central star(s) or blown out of the central regions by stellar winds.\\
\indent In Fig. 2, the CO emissions are well associated with the 8.0 $\mu$m emission; their morphologies are similar, especially  C$^{18}$O emission. The CO emissions are faint in the central regions of N4. Also Fig. 2 shows that the $^{12}$CO and $^{13}$CO molecular distribution is wider than PAHs distribution. There are some possible reasons. First, the critical density of C$^{18}$O is the highest, $^{13}$CO is lower and$^{12}$CO is the lowest.  $^{12}$CO traces diffused molecular gas, while C$^{18}$O traces dense molecular gas. Therefore the $^{12}$CO is distributed much wider than C$^{18}$O. Second, PAH emission originates from Photo Dissociation Regions (PDR). PAHs traces warm regions, while CO traces cold regions. Therefore the CO molecular gas is much more than PAHs.
\begin{figure}[!hbtp]
\centering
\includegraphics[width=0.3\textwidth]{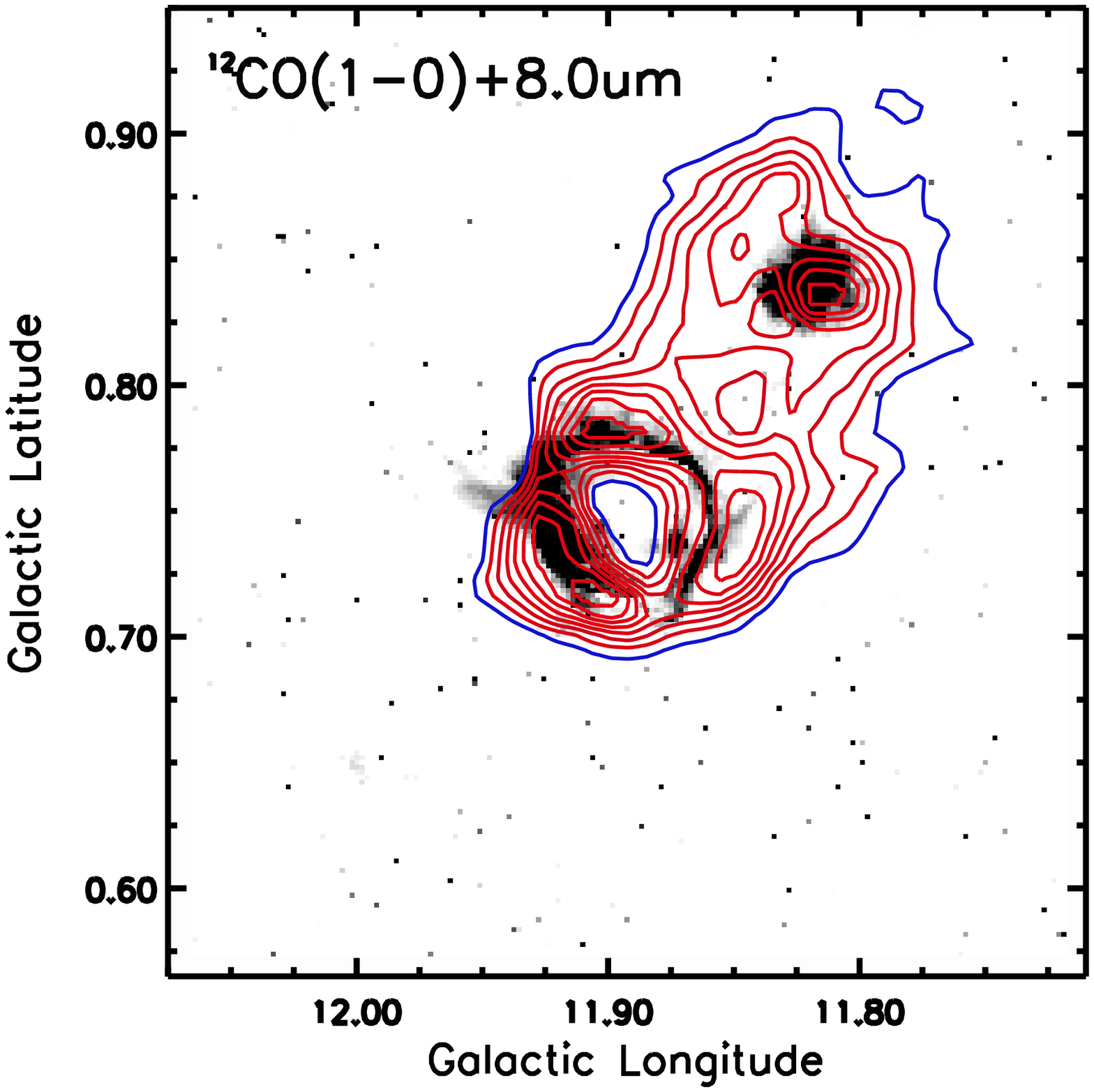}
\includegraphics[width=0.3\textwidth]{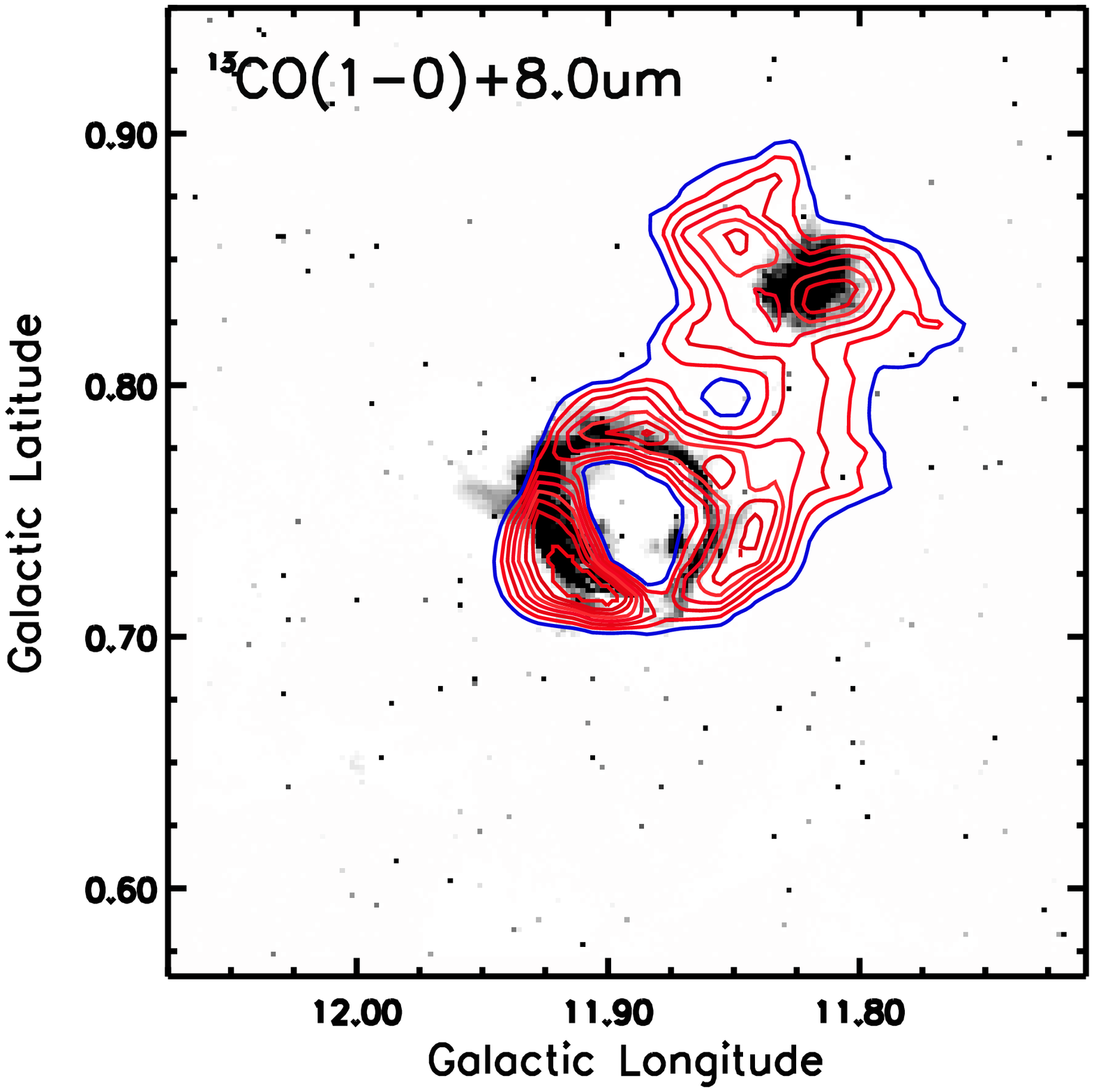}
\includegraphics[width=0.3\textwidth]{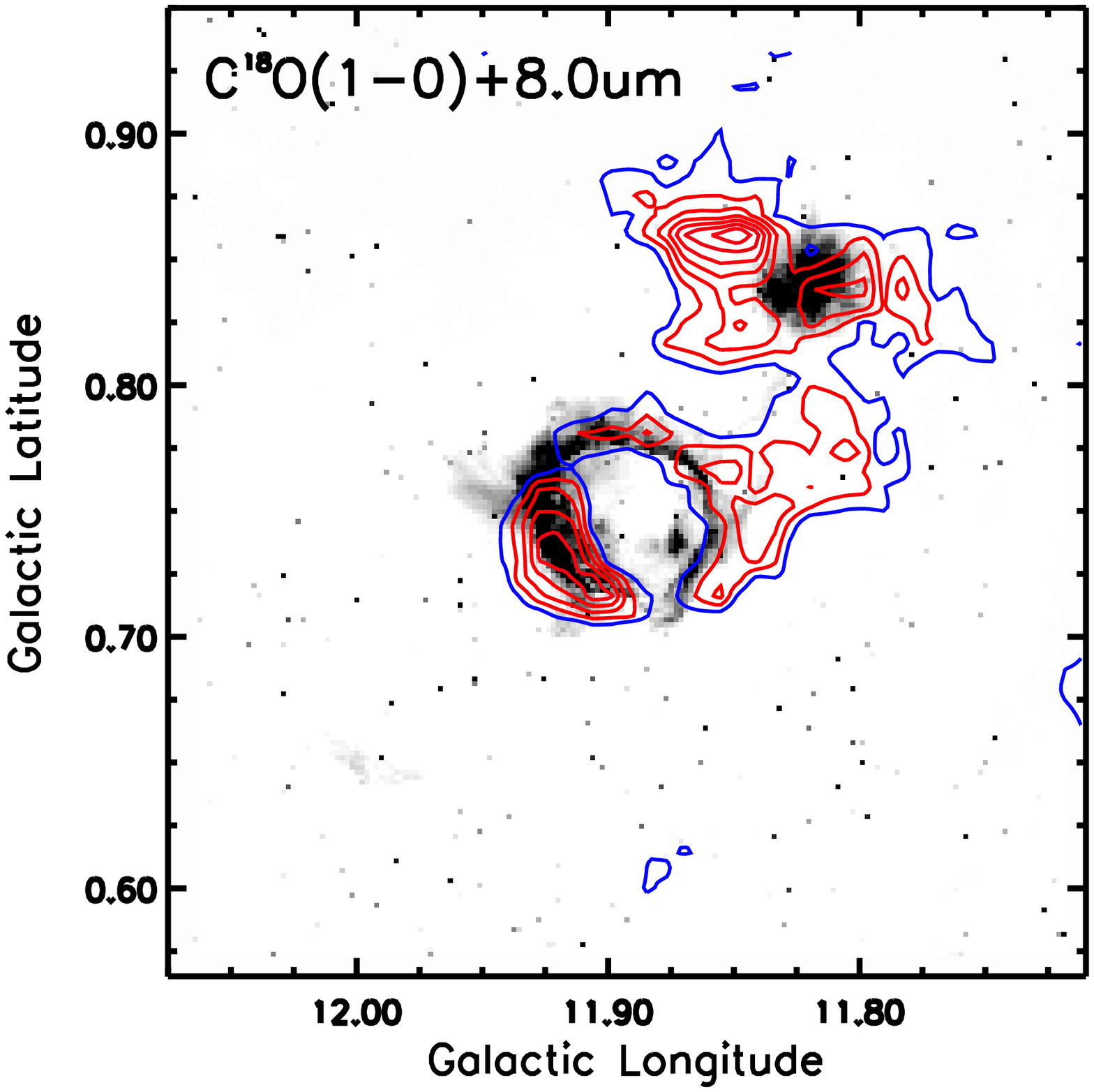}
\caption{Left panel: $^{12}$CO integrated intensity contours (for 21 km s$^{-1}$ $<$ $V_{LSR}$ $<$ 28 km s$^{-1}$) overlaid on Spitzer IRAC 8.0 $\mu$m image of N4 from the GLIMPSE. Contour levels start at 30$\%$ ($\sim$ 36 K km s$^{-1}$) of the peak $^{12}$CO integrated intensity, at a step $\sim$ 12 K km s$^{-1}$. Middle panel: $^{13}$CO integrated intensity contours (for 21 km s$^{-1}$ $<$ $V_{LSR}$ $<$ 28 km s$^{-1}$) overlaid on 8.0 $\mu$m image. Contour levels start at 30$\%$ ($\sim$ 12 K km s$^{-1}$) of the peak $^{13}$CO integrated intensity, at a step $\sim$ 4 K km s$^{-1}$. Right panel: C$^{18}$O integrated intensity contours (for 21 km s$^{-1}$ $<$ $V_{LSR}$ $<$ 28 km s$^{-1}$) overlaid on 8.0 $\mu$m image. Contour levels start at 30$\%$ ($\sim$ 1.65 K km s$^{-1}$) of the peak C$^{18}$O integrated intensity, at a step $\sim$ 0.55 K km s$^{-1}$.}
   \label{Fig2}
\end{figure}

\subsection{The velocity structure of N4}
More detailed information of emission is shown in Fig. 3. Each panel shows a map of $W_{CO}$ of $^{13}$CO emission in different velocity shown in the upper left corner. In the channel map, the $^{13}$CO emission of N4 first appear in the southwest direction, then enlarge as a semi-circle, a circle, again a semi-circle, at last disappear in the northeast direction. Fig. 3 indicates that the southwest part is blueshifted, while the northeast part is redshifted. As one steps through velocity space, a spherical shell would appear as a small blueshifted region (i.e., the front cap of the bubble), expanding into a ring (i.e., the midsection of the bubble), and then contracting back to a small redshifted region (i.e., the back cap of the bubble) (Arce et al. 2011). 
If bubbles are two dimensional projections of spherical shells, it can be expected to see the front and back faces of these shells at blueshifted and redshifted velocities (Beaumont $\&$ Williams 2010). Therefore if N4 is an expanding bubble, having spherical shape, the channel map of $^{13}$CO emission would appear different than shown in Fig. 3. However, with an inclined expanding ring, we would expect the observed channel map as shown in Fig. 3. In such a case, the front part of the inclined ring moves toward us (i.e., blue shift), and its back part moves away from us (i.e., redshift). Thus N4 is more likely a inclined expanding ring than a spherical bubble. \\
\indent There may exists other possibilities, for example, one most natural interpretation to draw from this observation is that the fronts and backs of N4 are missing, and we are instead observing only the bubble midsection. A corollary is that the molecular clouds in which N4 is embedded is oblate with thickness of a few parsecs, comparable to typical ring diameters. N4 expands and breaks out the molecular clouds along its axis and we see a ring of CO emission, approximately circular if the flattened axis is along our line of sight, otherwise elliptical. This kind situation may happen, as the density of molecular clouds are different at the front, middle and back parts of the bubble. Anyway we would like to believe that N4 has a ring structure, no matter it is actually a ring or a bubble that has no fronts and backs.

\begin{figure}[!htbp]
\centering
\includegraphics[width=0.9\textwidth]{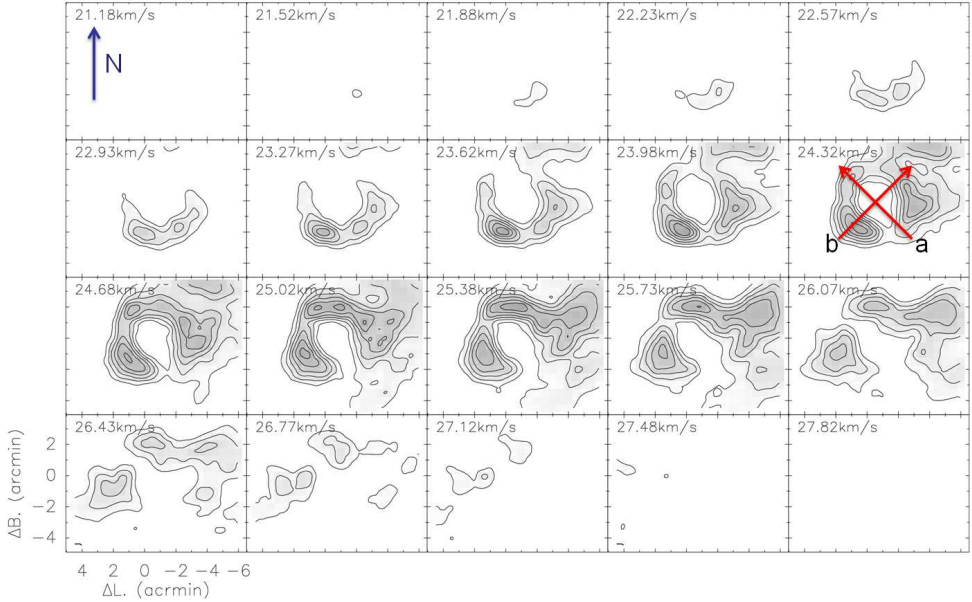}
\caption{The channel map of $W_{CO}$ integrated over two channels (for
21 km s$^{-1}$ $<$ $V_{LSR}$ $<$ 28 km s$^{-1}$). Contour levels start 40$\%$ ($\sim$ 16 K km s$^{-1}$) of the peak $^{13}$CO (1$-$0) integrated intensity, at a step $\sim$ 0.55 K km s$^{-1}$. 
The number on the upper left corner of each panel indicates the central $V_{LSR}$ of the channel map. The coordinate is relative to the center of N4 (11.892$^\circ$, 0.748$^\circ$).  Arrow a and b are labeled as the directions we make velocity-position map.}
   \label{Fig3}
\end{figure}

\subsection{The kinematics of N4 in $^{13}$CO (1$-$0) emission}
In order to study more detailed information about the kinematics of N4, we make the velocity-position map of the $^{13}$CO emission. We take two directions which are labeled as a (i.e., the northeast direction) and b (i.e., the northwest direction) shown in Fig. 3 to make the velocity-position map. Direction a is taken from the southwest point (the smallest velocity) to the center of N4 and extended to the outer boundary of N4; direction b is taken perpendicularly to direction a. Fig. 4 shows the velocity-position map of the $^{13}$CO emission of N4. The two red lines plotted on the two panels present the velocity of the molecular clouds in the vicinity of N4, which with a value of $\sim$ 24.3 km s$^{-1}$ is used as an indicator of system velocity of the local molecular clouds. Also we draw six points in each panel to show the central velocity of the inner, the middle and the outer point along each cut (a, b). The diagonal arrow plotted in panel (a) shows a slope trend that the velocity of southwest part of N4 is smaller than the northeast part, i.e., N4 has blue shift and red shift in direction a. Two panels also shows that the velocity of the inner region (blue point) is smaller than the velocity of the outer region (red point). \\
\indent To verify if all directions of N4 have same kind of property, we make a velocity-theta plot. Theta is the angle that starts from 0$^{\circ}$(direction b) and counts in counter-clockwise direction. We make three circles (the inner circle, the middle circle, the outer circle) and some radial lines start at the center of N4 with a step of 30$^{\circ}$, so each radial line has three crosses with three circle. We calculate the average velocity at crosses and name them as $V_{in}$,$V_{med}$,$V_{out}$, corresponding to the inner, middle, outer circle respectively. Fig. 5 shows the velocity-theta plot of the $^{13}$CO emission of N4.  In Fig. 5, the left panel shows $V_{med}$ at 0$^\circ$ $<$ $\theta$ $<$ 180$^\circ$ is larger than $V_{med}$ at 180$^\circ$ $<$ $\theta$ $<$ 360$^\circ$, suggesting the same conclusion with previous saying that the southwest part of N4 has blue shift and the northeast part of N4 has red shift. The right panel shows $V_{out}$$-$$V_{med}$ $>0$ (for 0$^\circ$ $<$ $\theta$ $<$ 360$^\circ$) and $V_{in}$$-$$V_{med}$ $<0$ (for 0$^\circ$ $<$ $\theta$ $<$ 360$^\circ$). This means that $V_{in}$ $<$ $V_{out}$ of N4 in all directions. This kind of phenomenon suggests N4 may has a roll-up motion. Because the resolution of the 13.7 m millimeter telescope is not good enough, we regard this kind of motion is marginally detected. Higher resolution observations are necessary to confirm this. 

\begin{figure}[!htbp]
\centering
\includegraphics[width=0.45\textwidth]{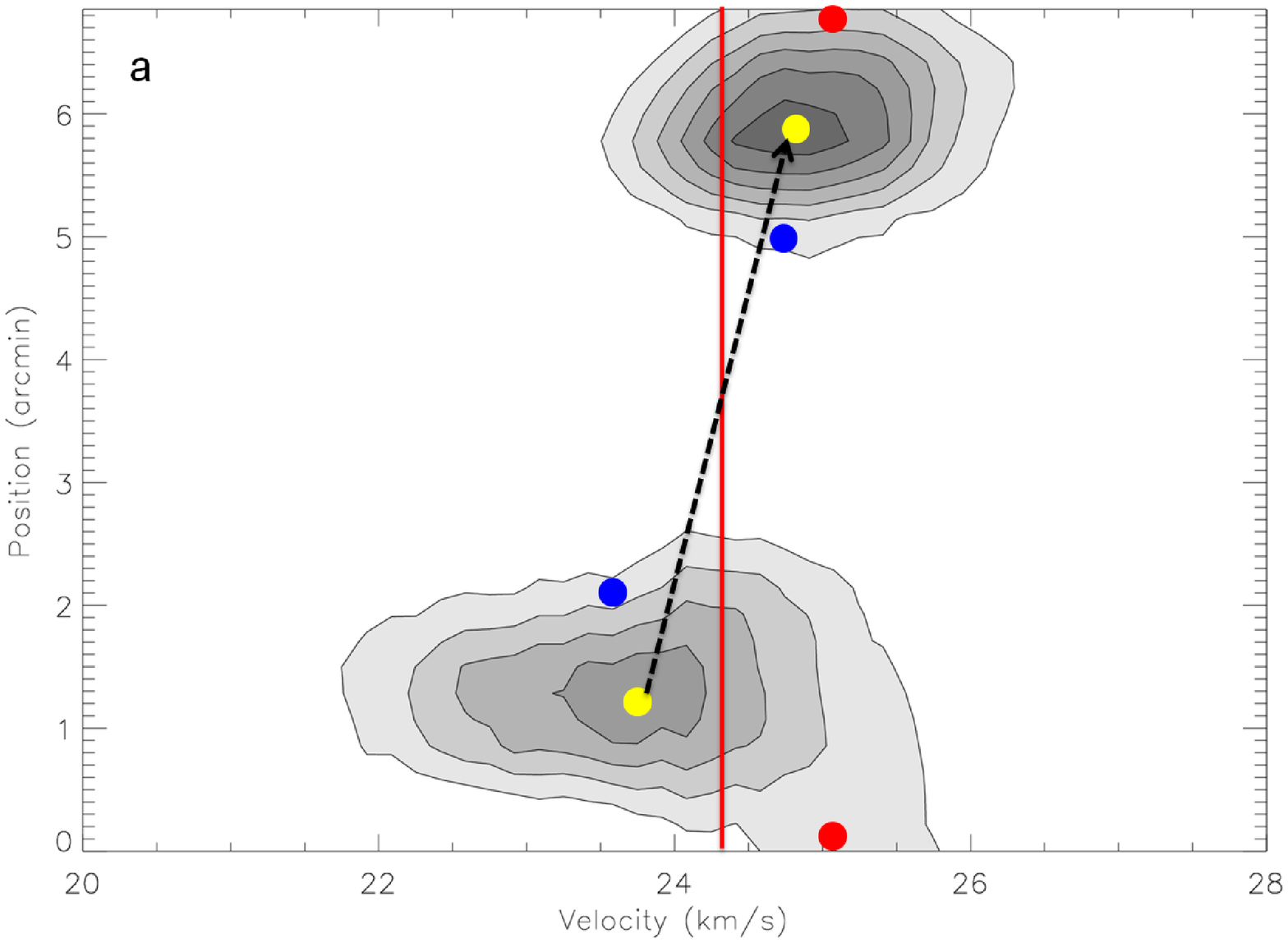}
\includegraphics[width=0.45\textwidth]{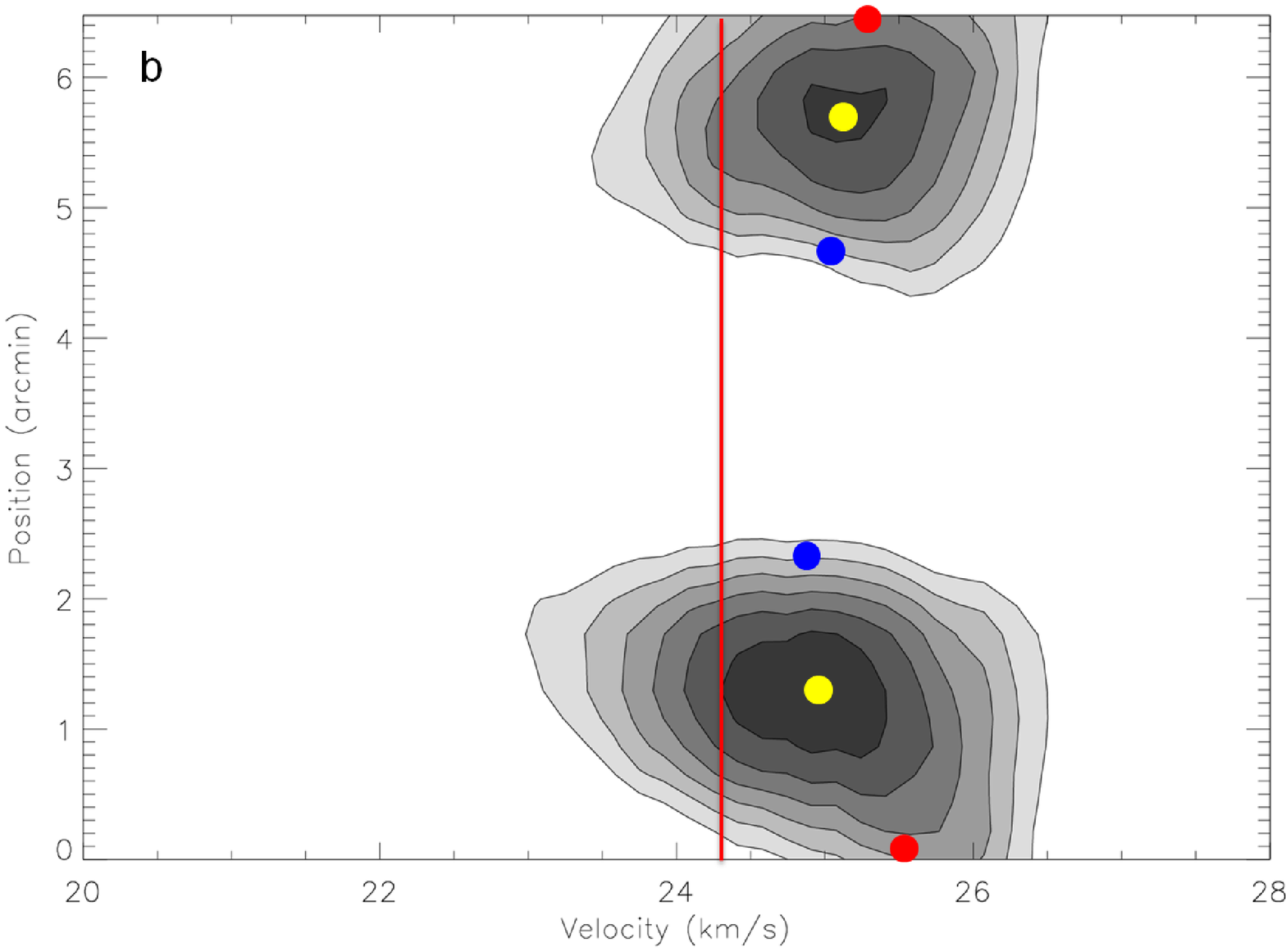}
\caption{The velocity-position map of the $^{13}$CO emission of N4. The two panels present two directions (a, b) taken which is a cross shown in Fig. 3. The two red lines plotted on the two panels present the velocity of the molecular clouds in the vicinity of N4, which is used as an indicator of system velocity of the local molecular clouds, the value is $\sim$ 24.3 km s$^{-1}$. Six points in each panel show the central velocity of the inner, the middle and the outer point along each cut (a, b). The contour levels are from 0.8 K to 4.2 K in steps of $\sim$ 0.6 K.}
   \label{Fig4}
\end{figure}

\subsection{Infall motion in N4}
Gravitational infall or core collapse can take place in high-mass young stellar objects (YSOs) at early stages and continue all the way to the stage of Ultra Compact (UC) H{\scriptsize II} regions (Keto 2003; Sollins, Ho $\&$ Paul 2005). Many infall candidates have been identified based on their spectral signature. They can serve as good targets to study the massive star birth and gas dynamics in the molecular cores (e.g. Ren et al. 2012). For the $^{12}$CO emission, the blueshifted emission peak is stronger than the redshifted one, and the central absorption dip is well coincident with the $^{13}$CO or C$^{18}$O line peak. Such blue asymmetric $^{12}$CO line suggest the presence of infall motion towards the core centre in the molecular cloud. When infall occurs in the envelope, where the outer region is cooler than the inner region, the gas in the front part (outer region) would absorb the redshifted emission, leaving a dip on the redshifted side (Zhou et al. 1993; Mardones et al. 1997).\\
\indent In N4, when we checked the spectra of the $^{12}$CO emission, we found one region shows this blue asymmetric signature. As shown in Fig. 6, we located the center position (L = 11.92$^\circ$, B = 0.74$^\circ$), then taken its nearby 20 (5$\times$4) positions to show their spectra in the $^{12}$CO, $^{13}$CO and C$^{18}$O. We can see there are 7 or 8 positions have the feature, i.e., blue asymmetric. Therefore it can be an infall candidate. However there are many components in the line of sight and the spectra are complex, and the resolution of 13.7 m telescope is not good enough, so we need higher resolution observations to confirm this infall candidate.

\begin{figure}[!htbp]
\centering
\includegraphics[width=12cm]{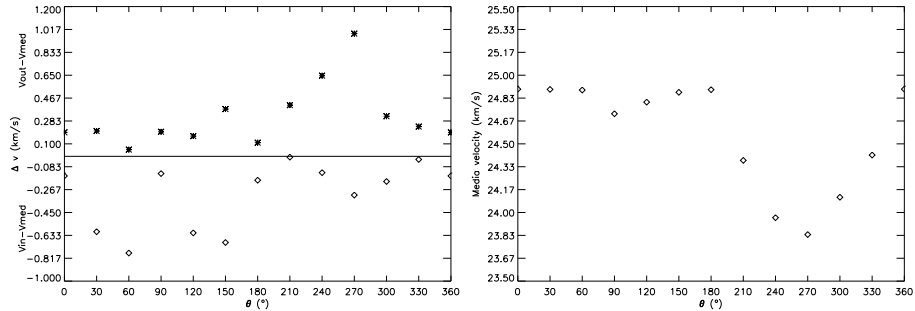}
\vspace{-7.5cm}
\caption{The velocity-theta plot of the $^{13}$CO emission of N4. $V_{in}$ is the velocity of the inner circle of N4, $V_{out}$ is the velocity of the outer circle of N4. $V_{med}$ is the velocity of the middle circle of N4.}
   \label{Fig5}
\end{figure}

\begin{figure}[!ht]
\centering
\includegraphics[angle=-90,width=0.8\textwidth]{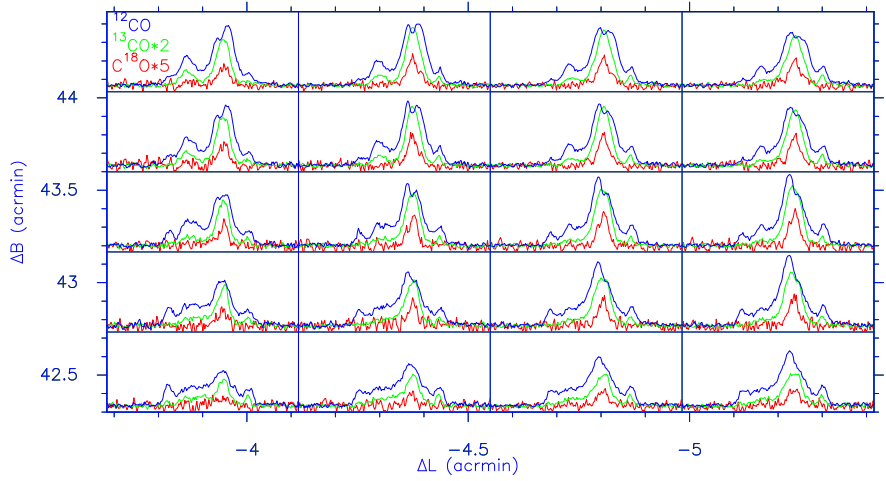}
\caption{The blue line is $^{12}$CO emission, the green line is $^{13}$CO emission and the red line is C$^{18}$O emission. The brightness temperature of each $^{13}$CO spectrum is multiplied by 2 and the brightness temperature of each C$^{18}$O spectrum is multiplied by 5. The coordinate is relative to L = 12.0$^\circ$ and B = 0$^\circ$.}
    \label{Fig7}
\end{figure}

\subsection{The spectra of the $^{12}$CO (1$-$0), $^{13}$CO (1$-$0) and C$^{18}$O (1$-$0)}
The total emission profiles of the $^{12}$CO, $^{13}$CO and C$^{18}$O, averaged over the entire region of N4 (the red circle) are shown in Fig. 7. The blue line represents the $^{12}$CO emission, green the $^{13}$CO emission and red the C$^{18}$O emission. We see the total emission of $^{12}$CO is multi-peaked, indicative of several components in the line of sight. The emission from N4 is the main peak with a velocity range of 21 km s$^{-1}$ $<$ $V_{LSR}$ $<$ 28 km s$^{-1}$. The peak brightness temperature $T_{R}^{*}$ of $^{12}$CO, $^{13}$CO and C$^{18}$O is $\sim$ 15.5 K, 5 K and 1 K, respectively.

\begin{figure}[!ht]
\centering
\includegraphics[width=0.7\textwidth]{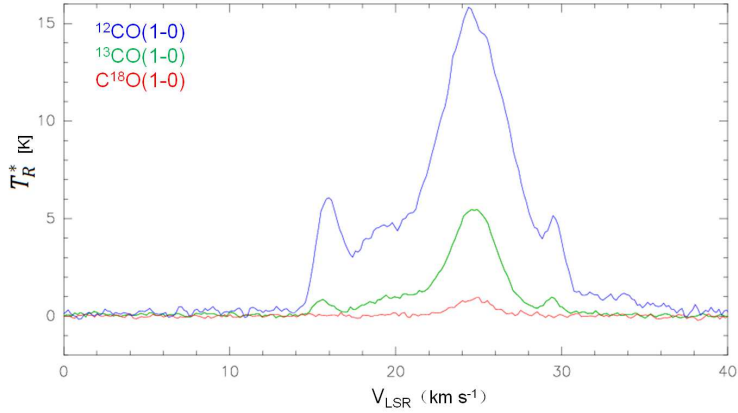}
\caption{The spectral lines of $^{12}$CO (1$-$0), $^{13}$CO (1$-$0) and C$^{18}$O (1$-$0), averaged over the entire region of N4.}
   \label{Fig1}
\end{figure}

\subsection{The physical parameters of N4}

By Gaussian fitting of the spectral lines of the $^{12}$CO, $^{13}$CO and C$^{18}$O, we obtained some parameters. As the spectral lines of N4 are multi-peaked (Fig. 7), the parameters of the main peak is adopted (see Table 1).

\begin{table}[h!!!]

\small
\centering
\begin{minipage}[]{100mm}
\caption[]{Fundamental parameters of CO emission}\label{Table 1}\end{minipage}
\tabcolsep 1mm
 \begin{tabular}{clcl}
  \hline\noalign{\smallskip}
$^{12}$CO (1$-$0)$\quad$            &    $\quad$$\quad$$\quad$$^{13}$CO (1$-$0) & C$^{18}$O (1$-$0)\\
  \hline\noalign{\smallskip}
$T_{R}^{*}(peak)$$\quad$     Width$\quad$$\quad$$\quad$      Vel$\quad$$\quad$ & $T_{R}^{*}(peak)$$\quad$    Width$\quad$$\quad$ $\quad$     Vel$\quad$&$T_{R}^{*}(peak)$$\quad$  Width$\quad$$\quad$      Vel$\quad$$\quad$\\
$\quad$$\quad$ (K)$\quad$$\quad$    (km s$^{-1}$)$\quad$ (km s$^{-1}$)$\quad$ &   $\quad$(K)$\quad$$\quad$$\quad$    (km s$^{-1}$)$\quad$ (km s$^{-1}$)$\quad$ &$\quad$ (K)$\quad$     $\quad$ (km s$^{-1}$)$\quad$    (km s$^{-1}$)\\                
  \noalign{\smallskip}\hline
15.1$\quad$$\quad$$\quad$ 5.1$\quad$$\quad$$\quad$   $\quad$24.7$\quad$  & $\quad$4.4$\quad$$\quad$$\quad$$\quad$     2.9$\quad$$\quad$$\quad$    24.7$\quad$ & $\quad$ 0.8$\quad$$\quad$$\quad$   2.8$\quad$$\quad$$\quad$   24.7$\quad$\\
\end{tabular}
\end{table}

Here we assume that the molecular clouds are in Local Thermodynamic Equilibrium (LTE) status and the line center of $^{12}$CO (1$-$0) is optically thick. Using the formula according to Rohlfs $\&$ Wilson 2000 and Nagahama et al. (1998), the physical parameters of N4 are calculated.
\begin{equation}
 \  T_{ex} = \frac{5.53}{ln(1+\frac{5.53}{T_{R}^{*}(^{12}CO)+0.82})} \ 
\label{eq:LebsequeI}
\end{equation}
here $T_{R}^{*} $ is the brightness temperature, then we can obtain the excitation temperature of $^{12}$CO (1$-$0), $T_{ex}$ $=$ 18.6 K. Assuming that $^{13}$CO (1$-$0) and C$^{18}$O (1$-$0) have the same excitation temperature as $^{12}$CO (1$-$0), the optical depth of $^{13}$CO and C$^{18}$O can be calculated using (2) $\&$ (3):

\begin{equation}
 \  \tau(^{13}CO)=-ln(1-\frac{T_{R}^{*}(^{13}CO)}{\frac{5.29}{exp(5.29/T_{ex}-1)}-0.89}) \ 
\label{eq:LebsequeI}
\end{equation}

\begin{equation}
 \  \tau(C^{18}O)=-ln(1-\frac{T_{R}^{*}(C^{18}O)}{\frac{5.27}{exp(5.27/T_{ex}-1)}-0.89}) \ 
\label{eq:LebsequeI}
\end{equation}

The column density of $^{13}$CO and C$^{18}$O can be obtained using (4) $\&$ (5):
\begin{equation}
 \  N(^{13}CO)=4.57\times 10^{13}(T_{ex}+0.89)exp(5.29/T_{ex})\int T_{R}^{*}(^{13}CO)dv \ 
\label{eq:LebsequeI}
\end{equation}

\begin{equation}
 \  N(C^{18}O)=4.57\times 10^{13}(T_{ex}+0.89)exp(5.27/T_{ex})\int T_{R}^{*}(C^{18}O)dv \ 
\label{eq:LebsequeI}
\end{equation}

 Assuming $N(H_{2})/N(^{13}CO)$ $=$ $7\times10^{5}$ (Nagahama et al. 1998) and $N(H_{2})/N(C^{18}O)$ $=$ $7\times10^{6}$ (Castets $\&$ Langer 1995), the average column density of hydrogen molecule and  the mass of the molecular cloud can be obtained, i.e., (6):

\begin{equation}
\ M=\mu m_{H} \sum (N(H_{2})_{i}\cdot A_{i})/2\times10^{30} \quad (M_{_\odot}) \
\end{equation}
here $\mu$ $=$ 2.8 is the mean molecular weight, $m_{H}$ is the mass of hydrogen atom and $A_{i}$ is the projected area of $^{13}$CO or C$^{18}$O emission region. Detailed information is shown in Table 2.\\
\begin{table}[h!!!]
\small
\centering
\begin{minipage}[]{100mm}
\caption[]{Derived physical parameters of N4}\label{Table 2}\end{minipage}
\tabcolsep 4mm
 \begin{tabular}{clcl}
 \hline\noalign{\smallskip}
 
Parameter&$^{13}$CO&C$^{18}$O\\
\\\hline
$A_{i}$ ($pc^{2}$)& $\quad$32.4&10.2 \\
$T_{ex}$ (K)&$\quad$18.6&18.6\\
$\tau$&$\quad$0.35&0.058\\
N(CO) ($cm^{-2}$)&$1.6\times10^{16}$&$3.0\times10^{15}$ \\
N$(H_{2})$ ($cm^{-2}$)&$1.1\times10^{22}$&$2.1\times10^{22}$ \\
M$(M_{_\odot})$&7.0 $\times 10^{3}$&4.3 $\times 10^{3}$\\
\hline\noalign{\smallskip}
\end{tabular}
\end{table}

As C$^{18}$O traces denser molecular gas than $^{13}$CO, the $A_{i}$ (C$^{18}$O) is smaller than $A_{i}$ ($^{13}$CO). The C$^{18}$O is optically thin, while $^{13}$CO is optically thick relatively to C$^{18}$O. Therefore the mass of N4 derived from the $^{13}$CO emission is larger than the mass derived from the C$^{18}$O emission at the assumed distance d = 3.2 kpc. Thus we suggests the mass of N4, i.e., $\sim$ 7.0$\times10^{3}$ $M_{_\odot}$ is more proper. Deharveng et al. (2010) has noted that N4 is surrounded by a shell of collected material, its mass is about 1150 $M_{_\odot}$ at d = 3.14 kpc. They use the 870 $\mu$m cold dust emission to estimate the amount of neutral material associated with N4. Because the dust continuum traces much more dense gas than $^{13}$CO and C$^{18}$O, the projected areas of $^{13}$CO and C$^{18}$O  emissions are larger than the 870 $\mu$m cold dust emission. Therefore it is reasonable that the mass of N4 we calculated is larger than that of Deharveng et al. (2010).

\begin{figure}[!hbtp]
\centering
\includegraphics[width=0.7\textwidth]{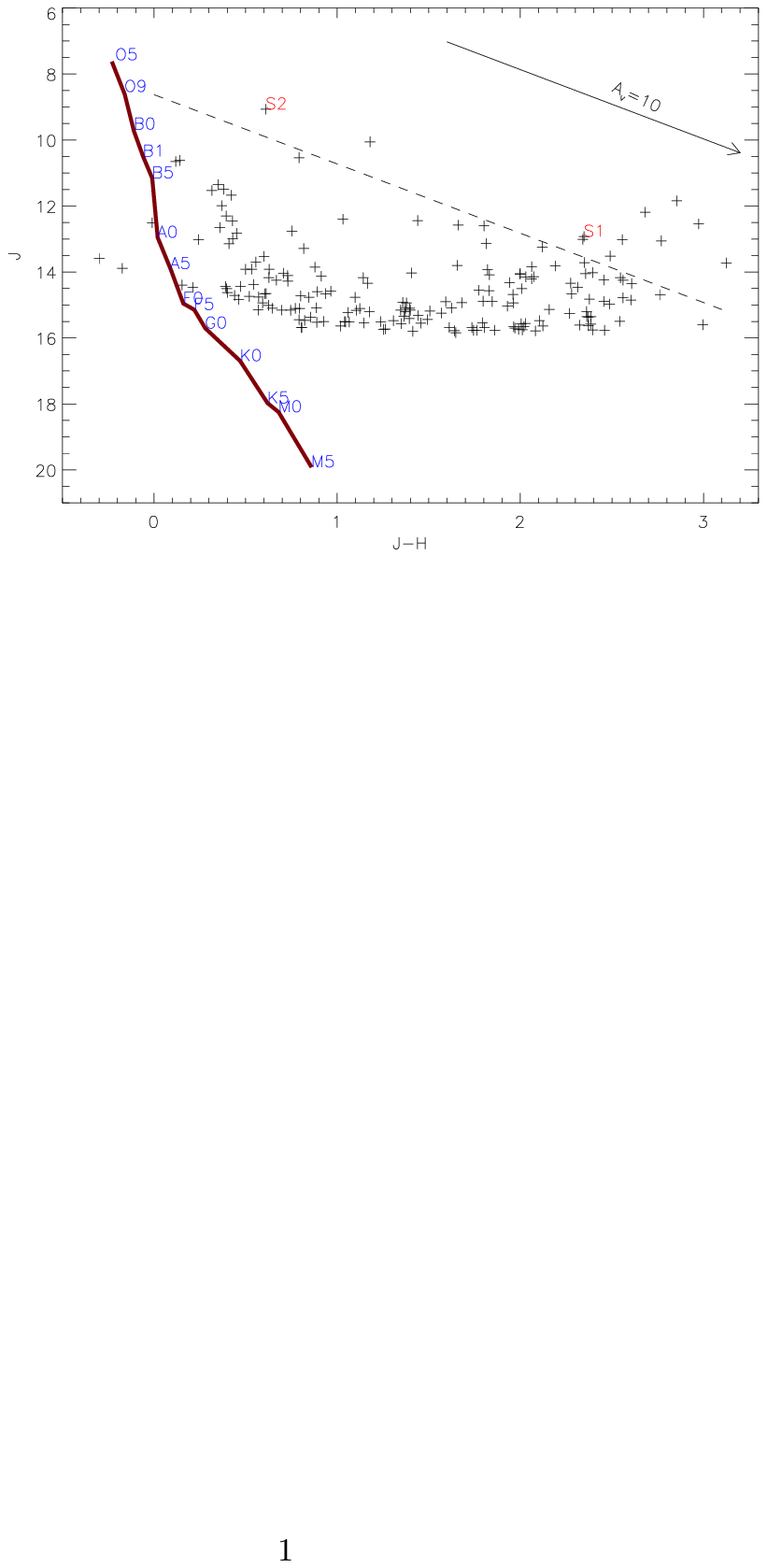}
\vspace{-14cm}
\caption{The (J, J$-$H) color-magnitude diagram for the stars in the N4 region. The solid line represents locus of the main sequence stars. The arrow plotted represents the extinction vector. The dashed line is parallel to the extinction vector and drawn from O9 main sequence.}
    \label{Fig8}
\end{figure}

\section{Discussion}
\label{sect:discussion}
\subsection{Energy source for N4}
The infrared dust bubble may be produced by exciting O- or B-type star(s), which are located inside the bubble. The ultraviolet (UV) radiation from exciting star(s) may heat dust and ionize the gas to form an expanding bubble shell (Watson et al. 2008). There are many stars inside and along the shell of N4. From the 2MASS (Skrutskie et al. 2006) point source catalogue, the (J, J$-$H) color-magnitude diagram is made, which is shown in Fig. 8. The solid line represents the locus of the main sequence stars (Covey et al. 2007). Several point sources which are located above the dashed line, we regard them as the massive star candidates. We find a massive star candidate near the center region of N4, labeled as S1, shown in Fig. 9. We compile the photometries available at different wavelengths (JHK + IRAC + WISE 12 $\mu$m + WISE 22 $\mu$m) for this candidate S1, see Fig. 10. Using standard models for circumstellar disks (e.g., Liu et al. 2012) and the well-tested radiative transfer code \texttt{MC3D} (Wolf 2003), we performed a SED fitting to derive the properties of this source. The results are shown in Fig. 10. The luminosity and the effective temperature of the best-fit model are ${\sim}25,000\,L_{\odot}$ and ${\sim}20,000\,\rm{K}$ respectively.
Interpolation of the pre-main sequence stellar models presented by Bernasconi (1996) suggests that S1 is a massive young stellar object with a mass of ${\sim}15\,M_{\odot}$ and an age of $\sim$ 1 Myr.  It correlates with the result shown in Fig. 8. A massive star (15 $M_{_\odot}$) with an age of a million year can produce energy $\sim$$2.6\times10^{44}$ J in UV band whereas the kinematic energy (E = 1/2 M$v^{2}$) of N4 is estimated $\sim$$1.4\times10^{39}$ J, implying that the UV photons produced by a massive star in 1 Myr is sufficient to drive the expansion of N4. Because of the position, age and energy output of S1, we think S1 may be the energy source for the expansion of N4, which deserves further studies.

\begin{figure}[!ht]
\centering
\resizebox{120mm}{!}{\includegraphics{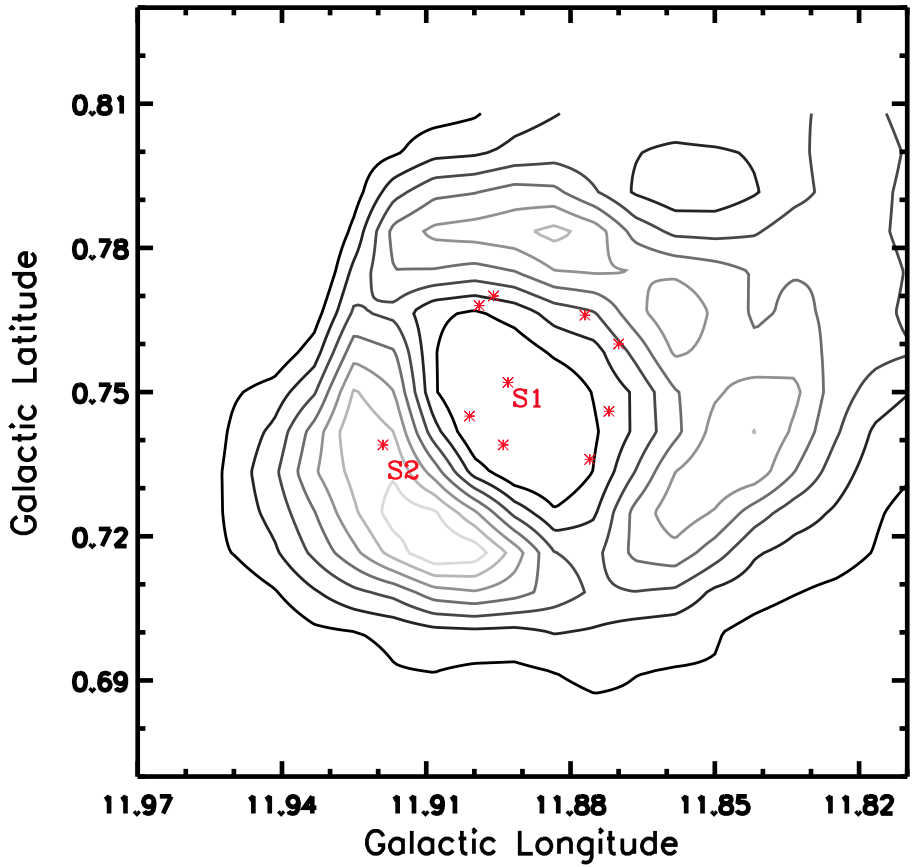}}
\caption{Overlay map of the massive stars positions on the$^{13}$CO (1$-$0) integrated intensity counters. Contour levels start at 30$\%$ ($\sim$ 12 K km s$^{-1}$) of the peak $^{13}$CO integrated intensity, at a step $\sim$ 4 K km s$^{-1}$. The red asterisks are labeled as the massive star candidates we found. S1 and S2 are labeled as the targets we discuss in the paper.}
    \label{Fig9}
\end{figure}

\subsection{Triggered star formation in N4}
Watson et al. (2010) had analyzed YSOs distribution around several mid-infrared-indentified bubbles including N4. Several YSOs are identified inside the shell, along the shell, and outside the shell, but there is no clear peak of YSO distribution in the N4 region. \\
They had not found any evidence of triggered star formation in N4. They claimed the reason may be that triggered star formation is difficult to identify there, either because of contamination in identifying YSOs or because the triggering mechanism does not dominate.\\
\indent However, we may have found the evidence of triggered star formation in N4. As mentioned before, there exists infall motion signature in N4, implying a high-mass YSO is forming. According to the velocity structure and the CO distribution of N4, we know N4 is expanding. This process can result interaction with the surrounding molecular clouds and trigger massive star formation near the shell clumps. Therefore we believe this forming YSO is triggered by the expansion of N4.\\
\indent As an interesting thing, a star labeled as S2 in Fig. 9 which is located above the dashed line in Fig. 8 is selected as a massive star candidate, whose coordinate (L = 11.92$^\circ$, B = 0.74$^\circ$) is consistent with the infall candidate we have found. It suggests that S2 and the infall candidate are probably the same source. Since S2 is embedded in N4, we think the formation process of S2 may be triggered by N4 expansion. As the resolution of our observation is not good enough, higher resolution observation is needed to confirm this. 

\begin{figure}[!htbp]
\begin{center}
\includegraphics[width=0.5\textwidth]{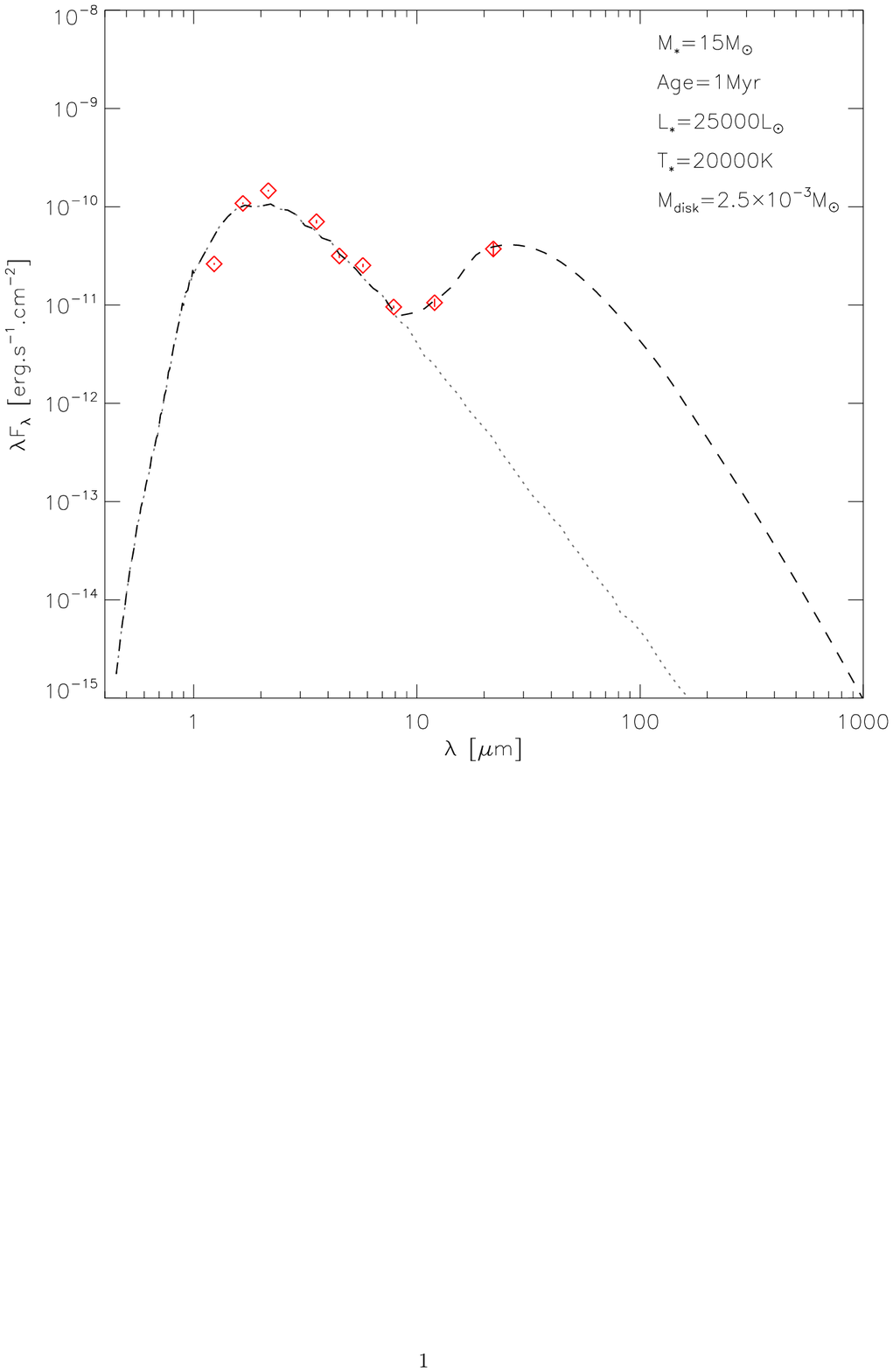}
\vspace{-5.cm}
\caption{The SED-fitting results of S1. The black dashed line represents the best-fit model, whereas the 
stellar contribution to the total flux of the system is depicted as a grey dotted line.
The observed data points are marked as diamonds (JHK + IRAC + WISE 12 $\mu$m + WISE 22 $\mu$m).}
    \label{Fig10}
\end{center}
\end{figure}

\section{Conclusions}
\label{sect:summary}
We have carried out simultaneous observations of the three CO (J = 1$-$0) ($^{12}$CO, $^{13}$CO and C$^{18}$O) line emissions towards N4. Reviewing the CO data of N4, in conjunction with the GLIMPSE data, enables us to draw following conclusions:\\
$1.$ The CO emissions are faint in the central regions of N4 and associated very well with the 8.0 $\mu$m emission; their morphologies are similar, especially the C$^{18}$O emission.\\
$2.$ The mass of N4 is $\sim$ 7.0$\times10^{3}$ $M_{_\odot}$ ($^{13}$CO). \\
$3.$ N4 is more likely an inclined expanding ring than a spherical bubble. \\
$4.$ There exists infall motion signature in N4, it can be a good infall candidate to study. Star formation may triggered by the expansion of N4. \\
$5.$ S1 may be the energy source for the expansion of N4.\\
\\

\normalem
\begin{acknowledgements}
We would like to thank the 13.7 m observatory staffs of Purple Mountain Observatory at the Qinghai Station for their supports during the observation and the supports from Millimeter $\&$ Sub-Millimeter Wave Laboratory of Purple Mountain Observatory. The GLMPSE data of this work is based on observations made with the Spitzer Space Telescope, which is operated by the Jet Propulsion Laboratory, California Institute of Technology under a contract with NASA. This publication makes use of data products from the Two Micron All Sky Survey, which is a joint project of the University of Massachusetts and the Infrared Processing and Analysis Center/California Institute of Technology, funded by the National Aeronautics and Space Administration and the National Science Foundation. The project is supported by the NSFC 10873037 and 10921063, China, and partially supported by the Ministry of Science and Technology (2007CB815406) of China.
\end{acknowledgements}



\label{lastpage}

\end{document}